\documentclass[journal=jacsat,manuscript=article]{achemso}
\usepackage[version=3]{mhchem}

\author{Daniel Willimetz}
\affiliation[Unknown University]
{Department of Physical and Macromolecular Chemistry, Charles University, Hlavova 8, Praha 2, 12800, Czech Republic}
\author{Lukáš Grajciar}
\email{lukas.grajciar@natur.cuni.cz}
\affiliation[Unknown University]
{Department of Physical and Macromolecular Chemistry, Charles University, Hlavova 8, Praha 2, 12800, Czech Republic}

\title[An \textsf{achemso} demo]
  {A Simple and Scalable Kernel Density Approach for Reliable Uncertainty Quantification in Atomistic Machine Learning}

\keywords{Machine learning, kernel density approximation, prediction uncertainty}

\begin{document}






\begin{abstract}

Machine learning models are increasingly used to predict material properties and accelerate atomistic simulations, but the reliability of their predictions depends on the representativeness of the training data. We present a scalable, GPU-accelerated uncertainty quantification framework based on $k$-nearest-neighbor kernel density estimation (KDE) in a PCA-reduced descriptor space. This method efficiently detects sparsely sampled regions in large, high-dimensional datasets and provides a transferable, model-agnostic uncertainty metric without requiring retraining costly model ensembles. The framework is validated across diverse case studies varying in: i) chemistry, ii) prediction models (including foundational neural network), iii) descriptors used for KDE estimation, and iv) properties whose uncertainty is sought. In all cases, the KDE-based score reliably flags extrapolative configurations, correlates well with conventional ensemble-based uncertainties, and highlights regions of reduced prediction trustworthiness. The approach offers a practical route for improving the interpretability, robustness, and deployment readiness of ML models in materials science.

\end{abstract}


Machine learning (ML) has become an indispensable tool in materials science due to its ability to efficiently model complex atomic interactions and material properties.\cite{Zhong2022Explainable} Among various approaches, neural network potentials (NNP) have gained widespread popularity for their speed and accuracy in predicting energies and forces.\cite{Duignan2024Potential} However, the reliability of NNP or other ML-based model predictions critically depends on the quality and representativeness of the training database. When test configurations fall outside the domain covered by training data, models extrapolate, which leads to large prediction errors.\cite{Kaser2023Neural} Accurate uncertainty estimation for predictions of NNPs and other models is therefore essential, motivating the development of methods aimed at quantifying the confidence of model prediction.\cite{Vandermause2020On,Jinnouchi2019Phase,sluijterman2023optimaltrainingmeanvariance,Zhu2022Fast,Heid2024Spatially,Bilbrey2025Uncertainty,Musielewicz2024Improved} A golden standard in uncertainty estimation is the ensemble or so-called "query-by-committee" method, which trains multiple models to capture prediction variance.\cite{Behler2007Generalized} While reasonably reliable, this approach is inefficient both at training and inference for large datasets as it requires training and deploying multiple models, making it computationally expensive.\cite{Tan2023Single}

Here, we present a single-model uncertainty quantification approach well scaling to large datasets based on a GPU-accelerated $k$-nearest-neighbor kernel density estimate (KDE) in the space of local atomic descriptors, with further efficiency gained by reducing their full dimension using principal component analysis (PCA).\cite{Mackiewicz1993Principal} First, descriptors are computed and stored for each atom present in the training dataset structures. Then, in inference, for example, during molecular simulation, the similarity of each encountered atomic descriptor (i.e., a query descriptor $\mathbf{q}$) to those present in the training set is efficiently evaluated using FAISS-based nearest-neighbor search.\cite{faiss} The similarity corresponds to a local density value $\rho$ for a query descriptor $\mathbf{q}$ given by
\begin{equation}
\rho_k(\mathbf{q}) = \frac{1}{k} \sum_{i \in \mathcal{N}_k(\mathbf{q})} \exp\!\left(-\frac{\lVert \mathbf{q} - \mathbf{x}_i\rVert^2}{2h^2}\right),
\label{eq:KDE_dens}
\end{equation}
where $\mathcal{N}_k(\mathbf{q})$ denotes the set of $k$ nearest training descriptors $\mathbf{x}_i$ and $h$ is the kernel bandwidth, taken as the standard deviation of distances within the training set. The kernel bandwidth, evaluated for each reference database separately, was calculated as the standard deviation of $k$ nearest-neighbor distances averaged over all reference database entries, which, as shown in Section S1 in the SI, outperforms common bandwidth estimators for proposed applications. 
The number of nearest neighbors, defined by the parameter $k$, was set to 100. As shown in Section S1, varying $k$ has only a minor effect on the results, indicating limited sensitivity to its exact value.

The advantage of our KDE-based approach is that unlike other single-model uncertainty methods,\cite{Zhu2022Fast,Nix1994Estimating,NEURIPS2020_aab08546} it can directly utilize general atomic descriptors without any additional training or fitting, beyond an automatic recalibration of kernel bandwidth for a reference database. Furthermore, these single-model techniques have been shown to underperform relative to the query by committee strategy.\cite{Tan2023Single} Our approach scales linearly with a very small prefactor (see below) to millions of atomic environments and provides an easily transferable uncertainty metric applicable to any local descriptor, such as SOAP\cite{Bart_k_2013} or descriptors from the MACE foundational model.\cite{batatia2024foundationmodelatomisticmaterials}. Lastly, the uncertainty in the approach presented here is evaluated at the level of atoms and their individual environments, which is more sensitive than the uncertainty estimates evaluated for whole (molecular) structures and provides a handle to focus the active learning strategies\cite{Roy2024Learning,Schwalbe2021Differentiable} to specific atoms and their environments. We note that a related strategy has been explored very recently,\cite{Schultz2025general} however, our GPU-accelerated $k$-nearest-neighbor KDE with PCA-reduced descriptors offers orders-of-magnitude lower computational cost. We demonstrate how the method performs in uncertainty estimation task in four case studies: 1) dynamics of platinum clusters on silica surfaces and in silica zeolites using both SchNet-based\cite{Schutt2017SchNet} and MACE-MP0\cite{batatia2024foundationmodelatomisticmaterials} descriptors, 2) water dynamics inside H-MFI zeolite using the MACE-MP0 foundation model descriptors,\cite{batatia2024foundationmodelatomisticmaterials} 3) the rMD17 benchmark using MACE-MP0 model and MACE model trained from scratch evaluating uncertainty for descriptors from both models, and 4) machine learning prediction of $^{27}$ Al NMR chemical shifts using SOAP-based\cite{Bart_k_2013} descriptors. All these tests are validated against literature data or ensemble-based estimates where available.

\begin{figure}[h!]
\centering
\includegraphics[width=0.5\linewidth]{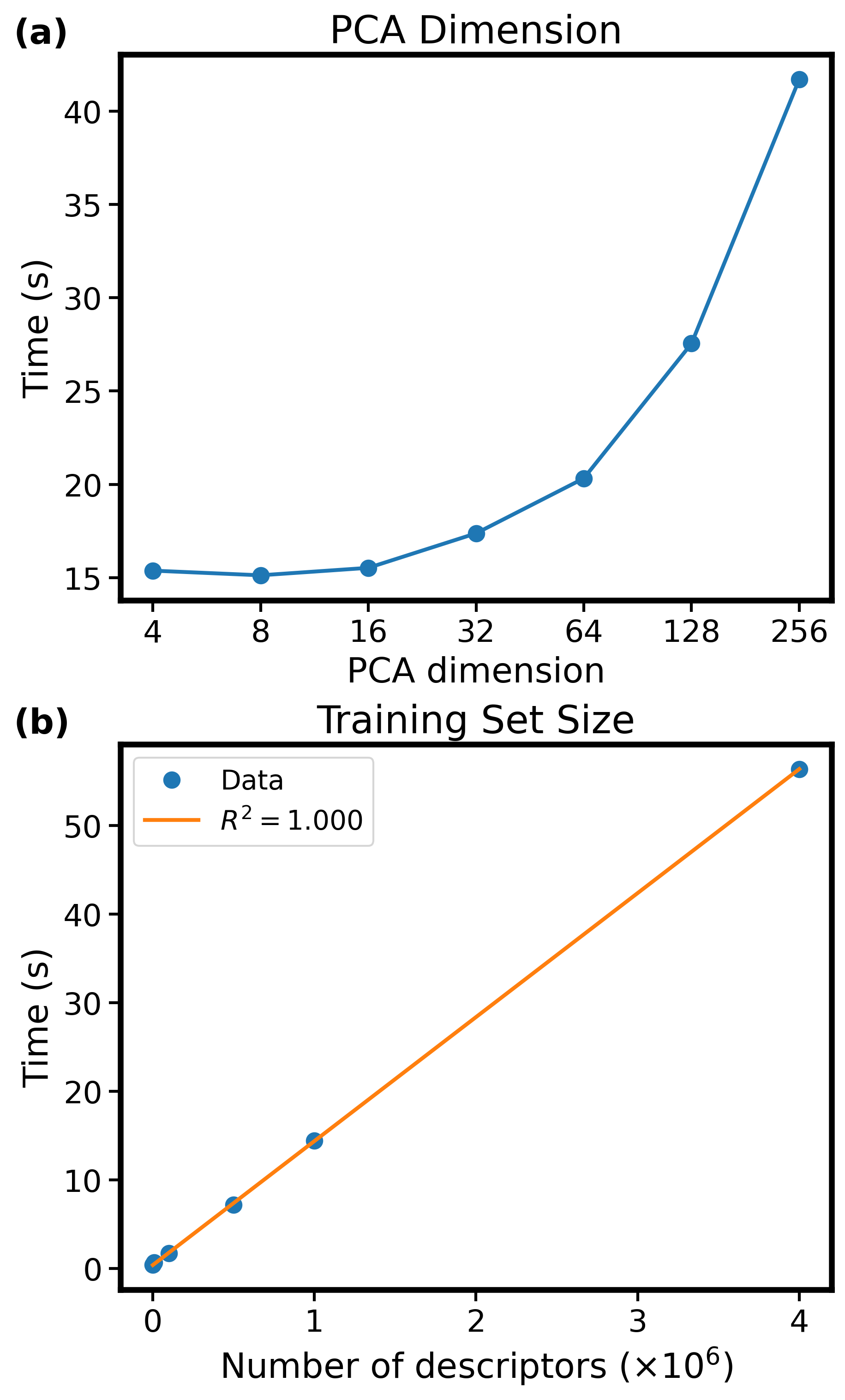}
\caption{
(a) Dependence of KDE-density evaluation time on the PCA-reduced dimension of the original atomic descriptor with the dimension of 256. The test is carried out for a fixed set of 228,000 atoms.
(b) Scaling of computational time with the number of atomic descriptors in the database for the 1,000 snapshots of the silicatene–Pt$_6$ system (containing 228,000 atoms in total with a 256-dimensional descriptor per atom).}
\label{fig:scaling}
\end{figure}

As a first application, we consider deploying the approach to evaluate the uncertainty of NNP prediction for platinum clusters in silicate environments, a system of catalytic relevance and large structural diversity. For this system a comprehensive database has been constructed and utilized in previous work.\cite{Benesova2025Mobility,Heard2024Migration} The reference dataset contains more than 230,000 structures. The ensemble of NNP models was trained on different random train/test splits of this dataset using the SchNet architecture,\cite{Schutt2017SchNet,Benesova2025Mobility} the prediction of which serves as a benchmark to test the KDE-based approach described here. The descriptors for the KDE approach were generated for the training database using the MACE-MP0 foundational model,\cite{batatia2024foundationmodelatomisticmaterials} although the SchNet descriptors can also be used, with the results for the KDE approach using SchNet descriptors shown in Figure~S7. The original MACE-MP0 atomic descriptors are 256-dimensional; however, the PCA dimensionality reduction analysis shows that the full descriptors (see Figure~S9) can be reduced to 16-dimensional descriptors (that is, the full descriptors projected on a 16-dimensional subset of principal components with the largest eigenvalues) without a loss in uncertainty prediction performance. Hence, in all subsequent applications of the MACE descriptors, the KDE estimation will work on top of their 16-dimensional projections. This upfront dimensionality reduction leads to approximately 2-3$\times$ cost reduction at inference time (Figure~\ref{fig:scaling}a), with the scaling being sublinear, likely reflecting the high efficiency of the GPU-accelerated FAISS implementation.\cite{faiss}

\begin{figure}[p!]
    \centering
    \includegraphics[width=0.9\linewidth]{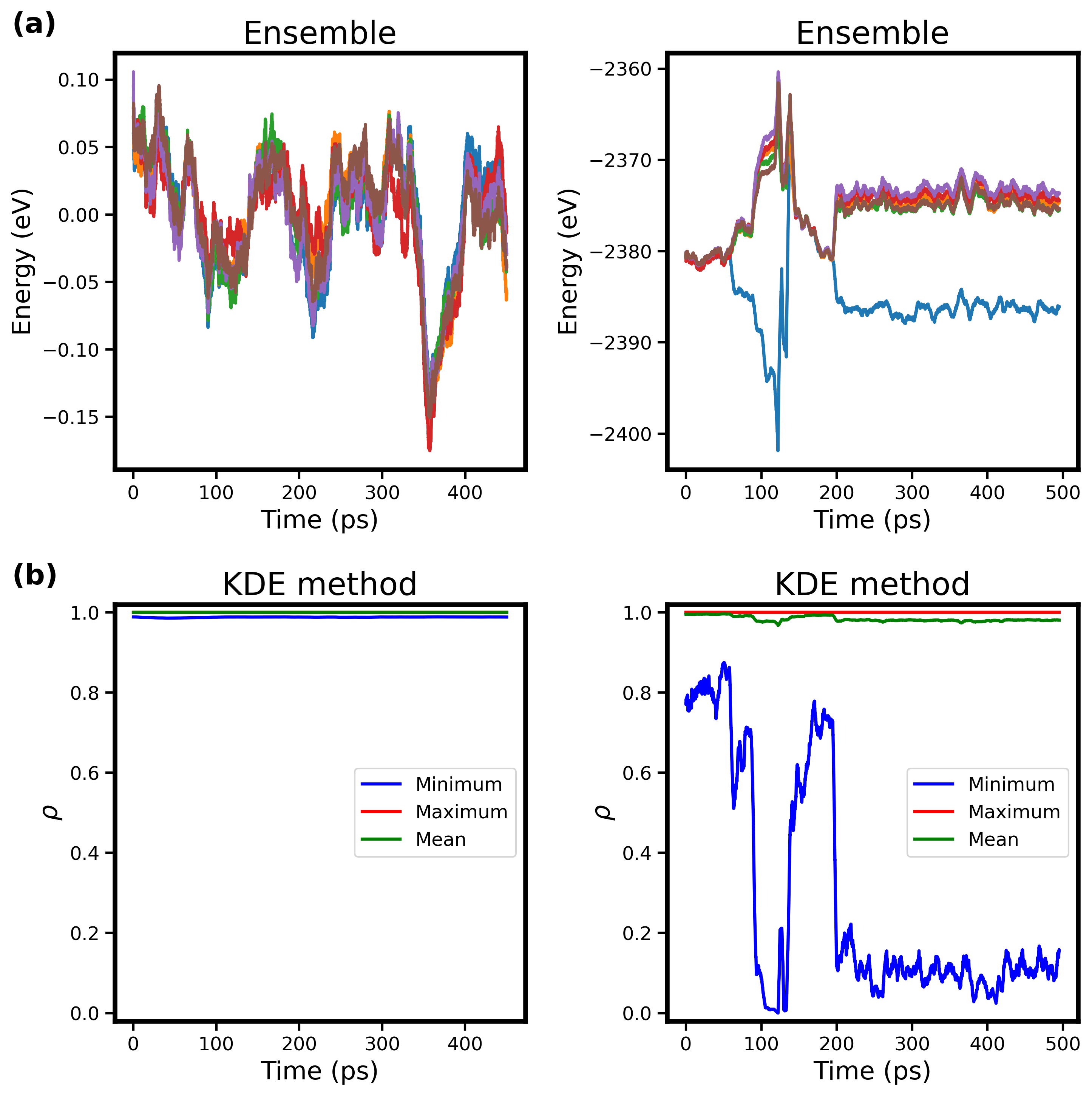}
    \caption{Comparison of the ensemble method (a) and the KDE method (b) for 500 ps MD simulations of Pt$_5$ clusters inside zeolite CHA at 750 K (left) and Pt$_6$ clusters on silicatene at 2000 K (right). The "mean", "minimum", and "maximum" data points represent the corresponding (mean, minimum and maximum) KDE densities evaluated for each atomic environment in each frame.}
    \label{fig:Pt_ens_KDE}
\end{figure}

Two types of platinum-on-silica systems were considered: (i) a Pt$_5$ cluster embedded inside zeolite CHA\cite{Heard2024Migration} and (ii) a Pt$_6$ cluster deposited on a defective silicatene layer (see Section S3 for more details).\cite{Benesova2025Mobility} The first case, Pt$_5$ inside CHA, is expected to be well represented in the training database, whereas Pt$_6$ on silicatene poses a greater challenge due to the under-representation of surfaces in the training database. For each case, molecular dynamics (MD) simulations were performed driven by a one model from the NNP ensemble, while the remaining five NNP models in the ensemble were used to assess potential extrapolation during the simulation.\cite{Heard2024Migration} Figure~\ref{fig:Pt_ens_KDE} then compares the (uncertainty) prediction from the benchmark NNP ensemble with the KDE-based estimate.

As expected, the KDE density of the structures sampled (i.e., of the query descriptor $\mathbf{q}$) for the Pt$_5$ system remains close to 1 throughout the MD simulation, indicating that it is well represented in the training set, in line with the prediction of the NNP ensemble that no extrapolation is expected. In contrast, the Pt$_6$ cluster on silicatene shows significant extrapolation, as one NNP model from the ensemble diverges from the predictions of the other models. This is mirrored by the prediction of the KDE method, where the density for the least similar atomic environments (i.e., "minimum" density) drops to almost zero at the same time that the NNP extrapolation occurs, indicating that these atomic environments are poorly represented in the training set. Although the average density remains near one, this shows that even a single atom can trigger extrapolation. Note also that only one of the NNP models in the ensemble indicates extrapolation, while others do not, which highlights the robustness of the KDE approach and indicates possible shortcomings and ambiguities of the ensemble method pertaining to the number of models in the ensemble and the way different models were obtained (different splits, random weight initialization, etc.).

Besides evaluating the performance of the KDE-method, we used the Pt$_6$-cluster-on-silicatene also as a case study to test the computational speed and scaling properties of the proposed method. One of the motivations behind this choice is the large size of the reference database,\cite{Benesova2025Mobility,Heard2024Migration} which contains over 4 million unique atomic environments, and which allows for a variable degree of subsampling to examine the scaling of KDE-method evaluation time with the database size. The local KDE density (Eq. \ref{eq:KDE_dens}) was evaluated for a test set of 1000 structures (composed of 228,000 atoms, that is, query descriptors $\mathbf{q}$) selected from the MD trajectory of Pt$_6$ cluster on silicatene.
Figure~\ref{fig:scaling}b shows the computational time of the KDE density as a function of the number of atoms in the training database. The results show clear linear scaling with a very small prefactor, underscoring both the efficiency of the method and its suitability for large systems and large datasets. For the largest training sets (4 million atoms) considered here, the KDE method completes in under one minute (i. e., 219 ms per 1000 test atoms on a single Tesla T4 GPU), compared to more than seven minutes for an ensemble of only five models. This advantage is even more pronounced for smaller training databases, where the KDE costs are dominated by the cost associated with the descriptor generation (see Section S4 for more details). But admittedly, for the KDE method one has to consider also an upfront computational cost due the evaluation of the training database representations, which is typically done once, with the representations saved and loaded at the inference time.

\begin{figure}[h!]
\centering
\includegraphics[width=0.5\linewidth]{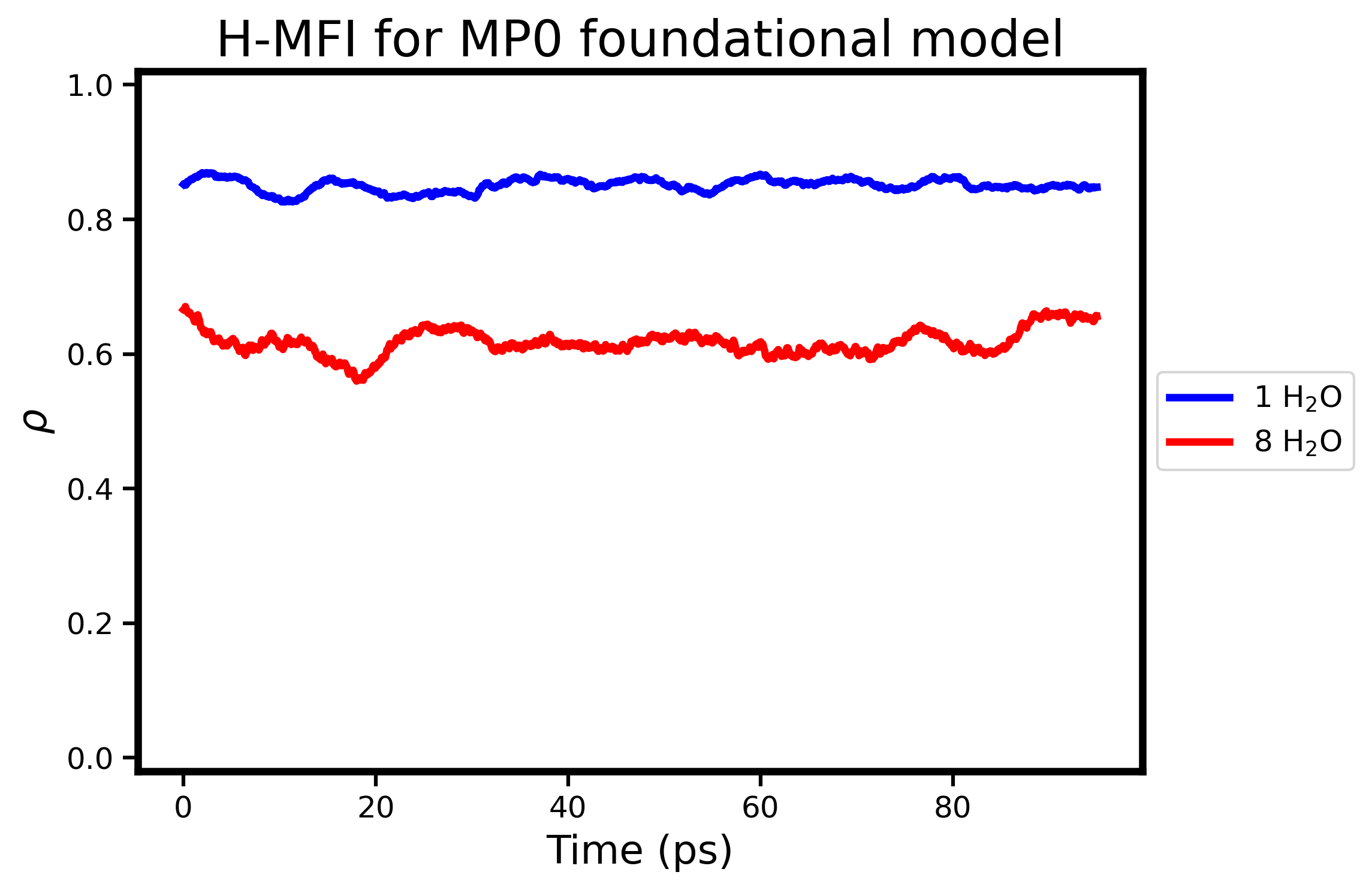}
\caption{The minimum KDE density of all atomic environments for 100 ps MD simulation of MFI zeolite at 350 K with 1 water per aluminium for Si/Al ratios of 95 (blue) and 11 (red).}
\label{fig:MP0}
\end{figure}

As a second case study, we evaluate the uncertainty of predictions for a foundational NNP model. The foundational models are becoming increasingly popular and widely used with an urgent need for a rigorous evaluation of their reliability. Here, we focus specifically on evaluating the reliability of the MACE-MP0 model\cite{batatia2024foundationmodelatomisticmaterials} to describe the aluminosilicate zeolite H-MFI (see Figure~S8) across varying aluminum and water loadings. In particular, two H-MFI systems were considered: (i) an H-MFI with a single aluminum (Si/Al = 95) and a single water molecule per unit cell, and (ii) an H-MFI with 8 aluminum (Si/Al = 11) and 8 water molecules per unit cell. These two systems are similar to those used in a recent study,\cite{Bilbrey2025Uncertainty} in which an ensemble of foundational models was generated to assess the prediction uncertainty of this popular foundational model. In both cases, all aluminum atoms were placed at the T5 site.\cite{IZA} The systems were subjected to 100 ps long equilibrium MD simulations using neural network potentials trained earlier on a comprehensive zeolite database\cite{Erlebach2024reactive} (see Section S3 for more details), and the resulting measures of uncertainties throughout the MD run are shown in Figure~\ref{fig:MP0}, with more details on the simulation set-up to be found in SI. The same KDE parameters (number of nearest neighbors considered, $k$ or kernel bandwidth $h$) were used as in the previous case study.

Figure~\ref{fig:MP0} shows that while the system with a small number of defects (and water) is fairly reliably described by the foundational model, the prediction uncertainty increases significantly for a higher water and aluminum content, as measured by the minimal local density for the atomic environments visited during MD runs. This appears to be consistent with both the fact that the MACE-MP0 database (i.e., MPTrj dataset\cite{deng_2023}) contains only pure silica and alumina systems (including zeolitic ones) but not defective systems, and with the recent observations by Bilbrey et al.\cite{Bilbrey2025Uncertainty} for a similar system using an ensemble of MACE-MP0 models (trained by fine-tuning the MACE-MP0 readout layer on different data splits and initializations). These findings highlight that the predictions from the off-the-shelf (MACE-MP0) foundational models\cite{batatia2024foundationmodelatomisticmaterials} should be approached with caution, particularly for systems that are under-represented in its training data.

\begin{figure}[h!]
\centering
\includegraphics[width=1\linewidth]{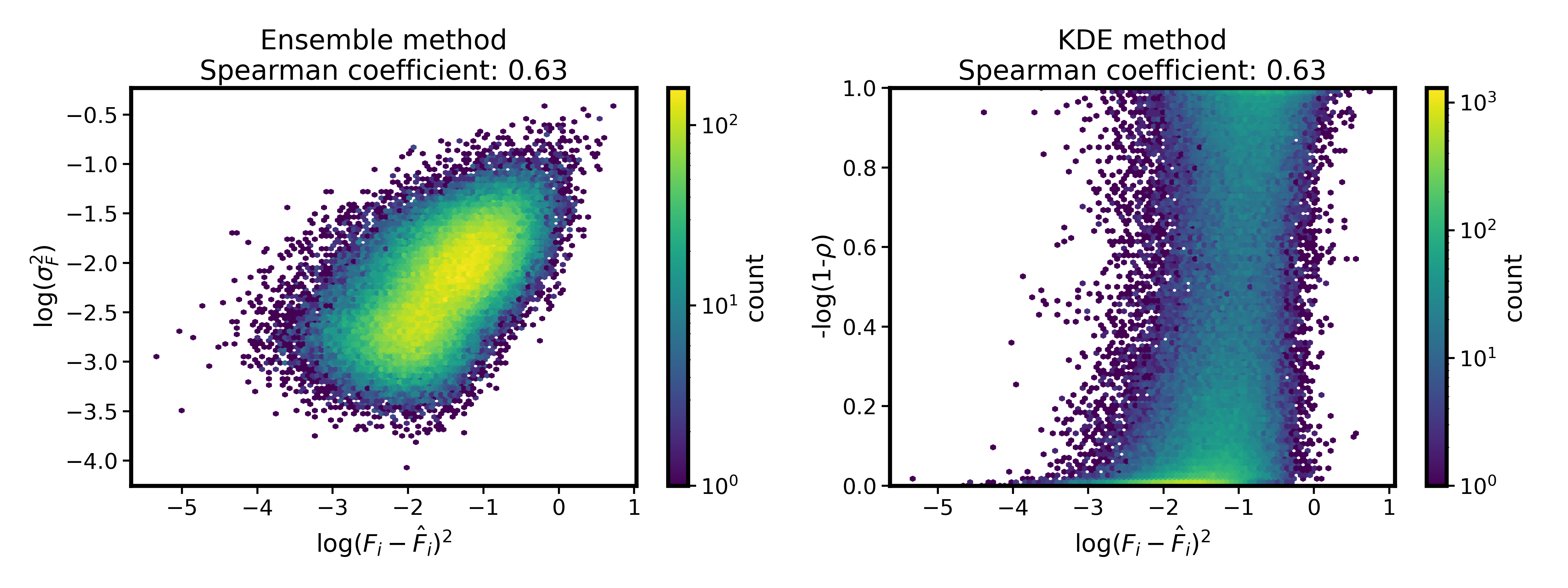}
\caption{Comparison of the ensemble method (left) and the KDE method (right) on the rMD17 dataset. $\hat{F}$ denotes the reference forces obtained from DFT calculations, and $\sigma_F^2$ represents the variance of the predicted forces from the neural network potentials in the ensemble.}
\label{fig:kde_ensemble}
\end{figure}

Next, we benchmark our approach on the rMD17 dataset,\cite{Christensen2020Revised} a widely used benchmark in atomistic machine learning. The dataset is comprised of short MD simulations of small organic molecules\cite{Christensen2020Revised} and provides specific train/test splits for training NNP ensembles, which we use here to train from scratch an ensemble of five MACE models (see Section S3 for more details).\cite{batatia2023macehigherorderequivariant} This dataset was extensively studied by Tan et al., who also reported how other uncertainty prediction methods perform for this dataset, making it an ideal test case for our method.\cite{Tan2023Single} Herein, the descriptors for the KDE approach were generated using the MACE-MP0 foundational model\cite{batatia2024foundationmodelatomisticmaterials} for the whole rMD17 dataset. Alternatively, we also considered obtaining descriptors from the MACE models trained from scratch, which show similar performance (Figure~S9) but we chose the MACE-MP0 representations for simplicity and transferability (i.e., without the need to train an NNP from scratch). The correlation between the force uncertainties predicted by the NNP ensemble and those obtained with our atom-based KDE density is shown in Figure~\ref{fig:kde_ensemble}.
Both methods yield the same Spearman correlation coefficient,\cite{Spearman} confirming the validity of the KDE-based approach. Moreover, for this benchmark, our approach outperforms the Gaussian Mixture Model,\cite{Zhu2022Fast,Tan2023Single} an alternative favorably scaling single-model approach for uncertainty estimation. In addition, the KDE method, in contrast to Gaussian Mixture Model does not require any additional (case-specific) training, enabling rapid evaluation of uncertainty using the dataset structures only. This is possible, in this case, by using general descriptors from the MACE-MP0 foundational model,\cite{batatia2024foundationmodelatomisticmaterials}, however, any other local structural descriptor could be used for this purpose, as shown in the last use case below.

To demonstrate the generalizability of our approach, we apply it in a very different setup, namely to estimate the uncertainty of $^{27}$Al chemical shift predictions using a kernel ridge regression (KRR) model\cite{Willimetz2025} built on top of the SOAP structural descriptors\cite{Bart_k_2013} and trained using a previously developed database\cite{Willimetz2025}. Although SOAP descriptors are high-dimensional, they can also be reduced to 16 dimensions, similarly to the MACE-based descriptors mentioned above, without significantly affecting the performance of the KDE-based approach in this case study (see Section S3 for more details). The KDE density is evaluated using these PCA-reduced SOAP descriptors of all the Al environments present in the reference database. The kernel bandwidth $h$ in this case is $\approx 30{,}000$. The specific systems in question are aluminosilicate zeolites with various topologies (CHA, MTT, and RTH), investigated in more detail in our previous studies.\cite{Willimetz2025,Willimetz2025Aluminum} In each framework, one silicon atom at the T1 position was replaced by an aluminum atom, a proton was added to balance the charge, and five water molecules were placed near the aluminium site. Subsequently, 1~ns molecular dynamics simulations were performed using neural network potentials
trained earlier on a comprehensive zeolite database,\cite{Erlebach2024reactive} and 50 structures were sampled for DFT NMR calculations using CASTEP (see SI for details).\cite{CASTEP} Table~\ref{tab:KRR} compares the chemical shieldings predicted by KRR and DFT, along with the mean absolute error (MAE) and the average KDE density score for each system.

\begin{table}[h!]
\centering
\caption{Comparison of DFT and KRR-predicted \(^{27}\)Al chemical shieldings (\(\delta\), ppm) for three zeolite frameworks, with mean absolute error (MAE) and average KDE density score using SOAP descriptors after PCA reduction to 16 dimensions.}
\label{tab:KRR}
\begin{tabular}{lcccc}
\hline
Zeolite & \(\delta\)(DFT) (ppm) & \(\delta\)(KRR) (ppm) & MAE & KDE density \\
\hline
CHA & 493.8 & 493.8 & 0.60 & 0.85 \\
MTT & 502.8 & 504.7 & 1.86 & 0.25 \\
RTH & 494.4 & 493.9 & 0.72 & 0.72 \\
\hline
\end{tabular}
\end{table}

The results show that the MAE between the KRR predictions and the DFT calculations correlates well with the average KDE density, indicating that the MTT framework is poorly represented in the training database. This aligns with our previous structural analysis,\cite{Willimetz2025Aluminum} where the high T–O–T angles in MTT were identified as a key factor contributing to its under-representation. Overall, these findings demonstrate that the KDE approach is broadly applicable across different types of atomic descriptors.

In this work, we have demonstrated that GPU-accelerated $k$-nearest-neighbor KDE offers a simple, yet surprisingly efficient and accurate, transferable approach to estimating uncertainty in atomistic machine learning models. Across multiple case studies, the method reliably identifies atomic environments under-represented in the training data, capturing extrapolation events with a sensitivity comparable to traditional ensemble approaches but at a fraction of the computational cost and without the need to train any additional models. Part of this performance likely stems from pragmatic heuristics that work well in practice, such as the automatic determination of KDE bandwidths, as well as the observation that despite very high-dimensional atomic representations, the intrinsic dimension of the datasets appears to be much lower, on the order of low tens (with 16 shown here to be sufficient). Although a deeper investigation of these aspects lies beyond the present scope, they may help explain why the approach performs robustly across diverse descriptors and ML architectures, including (foundational) neural network potentials and kernel ridge regression models. By providing a model-agnostic and computationally efficient metric of model confidence, this straightforward method enables more reliable predictions in materials simulations and chemical modeling.

\begin{acknowledgement}

The authors acknowledge the support of the Czech Science Foundation (23-07616S). This work was supported by the Ministry of Education, Youth and Sports of the Czech Republic through the e-INFRA CZ (ID:90254). In addition, Charles University Centre of Advanced Materials (CUCAM) (OP VVV Excellent Research Teams, project number \\
CZ.02.1.01/0.0/0.0/15\_003/0000417) is acknowledged. 

\end{acknowledgement}

\begin{suppinfo}

More details on the KDE approach, additional tests, and detailed description of used structures and databases are provided in the Supplementary Information. All Python scripts, molecular dynamics simulations, and setup files are available at DOI: 10.5281/zenodo.16855177

\end{suppinfo}

\bibliography{references}

\providecommand{\latin}[1]{#1}
\makeatletter
\providecommand{\doi}
  {\begingroup\let\do\@makeother\dospecials
  \catcode`\{=1 \catcode`\}=2 \doi@aux}
\providecommand{\doi@aux}[1]{\endgroup\texttt{#1}}
\makeatother
\providecommand*\mcitethebibliography{\thebibliography}
\csname @ifundefined\endcsname{endmcitethebibliography}  {\let\endmcitethebibliography\endthebibliography}{}
\begin{mcitethebibliography}{34}
\providecommand*\natexlab[1]{#1}
\providecommand*\mciteSetBstSublistMode[1]{}
\providecommand*\mciteSetBstMaxWidthForm[2]{}
\providecommand*\mciteBstWouldAddEndPuncttrue
  {\def\EndOfBibitem{\unskip.}}
\providecommand*\mciteBstWouldAddEndPunctfalse
  {\let\EndOfBibitem\relax}
\providecommand*\mciteSetBstMidEndSepPunct[3]{}
\providecommand*\mciteSetBstSublistLabelBeginEnd[3]{}
\providecommand*\EndOfBibitem{}
\mciteSetBstSublistMode{f}
\mciteSetBstMaxWidthForm{subitem}{(\alph{mcitesubitemcount})}
\mciteSetBstSublistLabelBeginEnd
  {\mcitemaxwidthsubitemform\space}
  {\relax}
  {\relax}

\bibitem[Zhong \latin{et~al.}(2022)Zhong, Gallagher, Liu, Kailkhura, Hiszpanski, and Han]{Zhong2022Explainable}
Zhong,~X.; Gallagher,~B.; Liu,~S.; Kailkhura,~B.; Hiszpanski,~A.; Han,~T. Y.-J. Explainable machine learning in materials science. \emph{npj Computational Materials} \textbf{2022}, \emph{8}, 204\relax
\mciteBstWouldAddEndPuncttrue
\mciteSetBstMidEndSepPunct{\mcitedefaultmidpunct}
{\mcitedefaultendpunct}{\mcitedefaultseppunct}\relax
\EndOfBibitem
\bibitem[Duignan(2024)]{Duignan2024Potential}
Duignan,~T.~T. The {Potential} of {Neural} {Network} {Potentials}. \emph{ACS Physical Chemistry Au} \textbf{2024}, \emph{4}, 232--241\relax
\mciteBstWouldAddEndPuncttrue
\mciteSetBstMidEndSepPunct{\mcitedefaultmidpunct}
{\mcitedefaultendpunct}{\mcitedefaultseppunct}\relax
\EndOfBibitem
\bibitem[K{\" a}ser \latin{et~al.}(2023)K{\" a}ser, Vazquez-Salazar, Meuwly, and T{\" o}pfer]{Kaser2023Neural}
K{\" a}ser,~S.; Vazquez-Salazar,~L.~I.; Meuwly,~M.; T{\" o}pfer,~K. Neural network potentials for chemistry: concepts, applications and prospects. \emph{Digital Discovery} \textbf{2023}, \emph{2}, 28--58\relax
\mciteBstWouldAddEndPuncttrue
\mciteSetBstMidEndSepPunct{\mcitedefaultmidpunct}
{\mcitedefaultendpunct}{\mcitedefaultseppunct}\relax
\EndOfBibitem
\bibitem[Vandermause \latin{et~al.}(2020)Vandermause, Torrisi, Batzner, Xie, Sun, Kolpak, and Kozinsky]{Vandermause2020On}
Vandermause,~J.; Torrisi,~S.~B.; Batzner,~S.; Xie,~Y.; Sun,~L.; Kolpak,~A.~M.; Kozinsky,~B. On-the-fly active learning of interpretable {Bayesian} force fields for atomistic rare events. \emph{npj Computational Materials} \textbf{2020}, \emph{6}, 20\relax
\mciteBstWouldAddEndPuncttrue
\mciteSetBstMidEndSepPunct{\mcitedefaultmidpunct}
{\mcitedefaultendpunct}{\mcitedefaultseppunct}\relax
\EndOfBibitem
\bibitem[Jinnouchi \latin{et~al.}(2019)Jinnouchi, Lahnsteiner, Karsai, Kresse, and Bokdam]{Jinnouchi2019Phase}
Jinnouchi,~R.; Lahnsteiner,~J.; Karsai,~F.; Kresse,~G.; Bokdam,~M. Phase {Transitions} of {Hybrid} {Perovskites} {Simulated} by {Machine}-{Learning} {Force} {Fields} {Trained} on the {Fly} with {Bayesian} {Inference}. \emph{Physical Review Letters} \textbf{2019}, \emph{122}, 225701\relax
\mciteBstWouldAddEndPuncttrue
\mciteSetBstMidEndSepPunct{\mcitedefaultmidpunct}
{\mcitedefaultendpunct}{\mcitedefaultseppunct}\relax
\EndOfBibitem
\bibitem[Sluijterman \latin{et~al.}(2023)Sluijterman, Cator, and Heskes]{sluijterman2023optimaltrainingmeanvariance}
Sluijterman,~L.; Cator,~E.; Heskes,~T. Optimal Training of Mean Variance Estimation Neural Networks. \emph{arXiv} \textbf{2023}, DOI: arXiv:2302.08875\relax
\mciteBstWouldAddEndPuncttrue
\mciteSetBstMidEndSepPunct{\mcitedefaultmidpunct}
{\mcitedefaultendpunct}{\mcitedefaultseppunct}\relax
\EndOfBibitem
\bibitem[Zhu \latin{et~al.}(2023)Zhu, Batzner, Musaelian, and Kozinsky]{Zhu2022Fast}
Zhu,~A.; Batzner,~S.; Musaelian,~A.; Kozinsky,~B. Fast uncertainty estimates in deep learning interatomic potentials. \emph{The Journal of Chemical Physics} \textbf{2023}, \emph{158}, 164111\relax
\mciteBstWouldAddEndPuncttrue
\mciteSetBstMidEndSepPunct{\mcitedefaultmidpunct}
{\mcitedefaultendpunct}{\mcitedefaultseppunct}\relax
\EndOfBibitem
\bibitem[Heid \latin{et~al.}(2024)Heid, Sch{\" o}rghuber, Wanzenb{\" o}ck, and Madsen]{Heid2024Spatially}
Heid,~E.; Sch{\" o}rghuber,~J.; Wanzenb{\" o}ck,~R.; Madsen,~G. K.~H. Spatially {Resolved} {Uncertainties} for {Machine} {Learning} {Potentials}. \emph{Journal of Chemical Information and Modeling} \textbf{2024}, \emph{64}, 6377--6387\relax
\mciteBstWouldAddEndPuncttrue
\mciteSetBstMidEndSepPunct{\mcitedefaultmidpunct}
{\mcitedefaultendpunct}{\mcitedefaultseppunct}\relax
\EndOfBibitem
\bibitem[Bilbrey \latin{et~al.}(2025)Bilbrey, Firoz, Lee, and Choudhury]{Bilbrey2025Uncertainty}
Bilbrey,~J.~A.; Firoz,~J.~S.; Lee,~M.-S.; Choudhury,~S. Uncertainty quantification for neural network potential foundation models. \emph{npj Computational Materials} \textbf{2025}, \emph{11}, 109\relax
\mciteBstWouldAddEndPuncttrue
\mciteSetBstMidEndSepPunct{\mcitedefaultmidpunct}
{\mcitedefaultendpunct}{\mcitedefaultseppunct}\relax
\EndOfBibitem
\bibitem[Musielewicz \latin{et~al.}(2024)Musielewicz, Lan, Uyttendaele, and Kitchin]{Musielewicz2024Improved}
Musielewicz,~J.; Lan,~J.; Uyttendaele,~M.; Kitchin,~J.~R. Improved {Uncertainty} {Estimation} of {Graph} {Neural} {Network} {Potentials} {Using} {Engineered} {Latent} {Space} {Distances}. \emph{The Journal of Physical Chemistry C} \textbf{2024}, \emph{128}, 20799--20810\relax
\mciteBstWouldAddEndPuncttrue
\mciteSetBstMidEndSepPunct{\mcitedefaultmidpunct}
{\mcitedefaultendpunct}{\mcitedefaultseppunct}\relax
\EndOfBibitem
\bibitem[Behler and Parrinello(2007)Behler, and Parrinello]{Behler2007Generalized}
Behler,~J.; Parrinello,~M. Generalized {Neural}-{Network} {Representation} of {High}-{Dimensional} {Potential}-{Energy} {Surfaces}. \emph{Physical Review Letters} \textbf{2007}, \emph{98}, 146401\relax
\mciteBstWouldAddEndPuncttrue
\mciteSetBstMidEndSepPunct{\mcitedefaultmidpunct}
{\mcitedefaultendpunct}{\mcitedefaultseppunct}\relax
\EndOfBibitem
\bibitem[Tan \latin{et~al.}(2023)Tan, Urata, Goldman, Dietschreit, and G{\' o}mez-Bombarelli]{Tan2023Single}
Tan,~A.~R.; Urata,~S.; Goldman,~S.; Dietschreit,~J. C.~B.; G{\' o}mez-Bombarelli,~R. Single-model uncertainty quantification in neural network potentials does not consistently outperform model ensembles. \emph{npj Computational Materials} \textbf{2023}, \emph{9}, 109\relax
\mciteBstWouldAddEndPuncttrue
\mciteSetBstMidEndSepPunct{\mcitedefaultmidpunct}
{\mcitedefaultendpunct}{\mcitedefaultseppunct}\relax
\EndOfBibitem
\bibitem[Ma{\' c}kiewicz and Ratajczak(1993)Ma{\' c}kiewicz, and Ratajczak]{Mackiewicz1993Principal}
Ma{\' c}kiewicz,~A.; Ratajczak,~W. Principal components analysis ({PCA}). \emph{Computers \&amp; Geosciences} \textbf{1993}, \emph{19}, 303--342\relax
\mciteBstWouldAddEndPuncttrue
\mciteSetBstMidEndSepPunct{\mcitedefaultmidpunct}
{\mcitedefaultendpunct}{\mcitedefaultseppunct}\relax
\EndOfBibitem
\bibitem[Johnson \latin{et~al.}(2019)Johnson, Douze, and J{\'e}gou]{faiss}
Johnson,~J.; Douze,~M.; J{\'e}gou,~H. Billion-scale similarity search with {GPUs}. \emph{IEEE Transactions on Big Data} \textbf{2019}, \emph{7}, 535--547\relax
\mciteBstWouldAddEndPuncttrue
\mciteSetBstMidEndSepPunct{\mcitedefaultmidpunct}
{\mcitedefaultendpunct}{\mcitedefaultseppunct}\relax
\EndOfBibitem
\bibitem[Nix and Weigend(1994)Nix, and Weigend]{Nix1994Estimating}
Nix,~D.; Weigend,~A. Estimating the mean and variance of the target probability distribution. Proceedings of 1994 IEEE International Conference on Neural Networks (ICNN'94). 1994; pp 55--60 vol.1\relax
\mciteBstWouldAddEndPuncttrue
\mciteSetBstMidEndSepPunct{\mcitedefaultmidpunct}
{\mcitedefaultendpunct}{\mcitedefaultseppunct}\relax
\EndOfBibitem
\bibitem[Amini \latin{et~al.}(2020)Amini, Schwarting, Soleimany, and Rus]{NEURIPS2020_aab08546}
Amini,~A.; Schwarting,~W.; Soleimany,~A.; Rus,~D. Deep Evidential Regression. Advances in Neural Information Processing Systems. 2020; pp 14927--14937\relax
\mciteBstWouldAddEndPuncttrue
\mciteSetBstMidEndSepPunct{\mcitedefaultmidpunct}
{\mcitedefaultendpunct}{\mcitedefaultseppunct}\relax
\EndOfBibitem
\bibitem[Bartók \latin{et~al.}(2013)Bartók, Kondor, and Csányi]{Bart_k_2013}
Bartók,~A.~P.; Kondor,~R.; Csányi,~G. On representing chemical environments. \emph{Physical Review B} \textbf{2013}, \emph{87}, 184115\relax
\mciteBstWouldAddEndPuncttrue
\mciteSetBstMidEndSepPunct{\mcitedefaultmidpunct}
{\mcitedefaultendpunct}{\mcitedefaultseppunct}\relax
\EndOfBibitem
\bibitem[Batatia \latin{et~al.}(2024)Batatia, Benner, Chiang, Elena, Kovács, Riebesell, Advincula, Asta, Avaylon, Baldwin, Berger, Bernstein, Bhowmik, Blau, Cărare, Darby, De, Pia, Deringer, Elijošius, El-Machachi, Falcioni, Fako, Ferrari, Genreith-Schriever, George, Goodall, Grey, Grigorev, Han, Handley, Heenen, Hermansson, Holm, Jaafar, Hofmann, Jakob, Jung, Kapil, Kaplan, Karimitari, Kermode, Kroupa, Kullgren, Kuner, Kuryla, Liepuoniute, Margraf, Magdău, Michaelides, Moore, Naik, Niblett, Norwood, O'Neill, Ortner, Persson, Reuter, Rosen, Schaaf, Schran, Shi, Sivonxay, Stenczel, Svahn, Sutton, Swinburne, Tilly, van~der Oord, Varga-Umbrich, Vegge, Vondrák, Wang, Witt, Zills, and Csányi]{batatia2024foundationmodelatomisticmaterials}
Batatia,~I. \latin{et~al.}  A foundation model for atomistic materials chemistry. \emph{arXiv} \textbf{2024}, DOI: arxiv:2401.00096\relax
\mciteBstWouldAddEndPuncttrue
\mciteSetBstMidEndSepPunct{\mcitedefaultmidpunct}
{\mcitedefaultendpunct}{\mcitedefaultseppunct}\relax
\EndOfBibitem
\bibitem[Roy \latin{et~al.}(2024)Roy, D{\" u}rholt, Asche, Zipoli, and G{\' o}mez-Bombarelli]{Roy2024Learning}
Roy,~S.; D{\" u}rholt,~J.~P.; Asche,~T.~S.; Zipoli,~F.; G{\' o}mez-Bombarelli,~R. Learning a reactive potential for silica-water through uncertainty attribution. \emph{Nature Communications} \textbf{2024}, \emph{15}, 6030\relax
\mciteBstWouldAddEndPuncttrue
\mciteSetBstMidEndSepPunct{\mcitedefaultmidpunct}
{\mcitedefaultendpunct}{\mcitedefaultseppunct}\relax
\EndOfBibitem
\bibitem[Schwalbe-Koda \latin{et~al.}(2021)Schwalbe-Koda, Tan, and G{\' o}mez-Bombarelli]{Schwalbe2021Differentiable}
Schwalbe-Koda,~D.; Tan,~A.~R.; G{\' o}mez-Bombarelli,~R. Differentiable sampling of molecular geometries with uncertainty-based adversarial attacks. \emph{Nature Communications} \textbf{2021}, \emph{12}, 5104\relax
\mciteBstWouldAddEndPuncttrue
\mciteSetBstMidEndSepPunct{\mcitedefaultmidpunct}
{\mcitedefaultendpunct}{\mcitedefaultseppunct}\relax
\EndOfBibitem
\bibitem[Schultz \latin{et~al.}(2025)Schultz, Wang, Jacobs, and Morgan]{Schultz2025general}
Schultz,~L.~E.; Wang,~Y.; Jacobs,~R.; Morgan,~D. A general approach for determining applicability domain of machine learning models. \emph{npj Computational Materials} \textbf{2025}, \emph{11}, 95\relax
\mciteBstWouldAddEndPuncttrue
\mciteSetBstMidEndSepPunct{\mcitedefaultmidpunct}
{\mcitedefaultendpunct}{\mcitedefaultseppunct}\relax
\EndOfBibitem
\bibitem[Sch{\" u}tt \latin{et~al.}(2018)Sch{\" u}tt, Kessel, Gastegger, Nicoli, Tkatchenko, and M{\" u}ller]{Schutt2017SchNet}
Sch{\" u}tt,~K.~T.; Kessel,~P.; Gastegger,~M.; Nicoli,~K.~A.; Tkatchenko,~A.; M{\" u}ller,~K.-R. SchNetPack: A {Deep} {Learning} {Toolbox} {For} {Atomistic} {Systems}. \emph{Journal of Chemical Theory and Computation} \textbf{2018}, \emph{15}, 448--455\relax
\mciteBstWouldAddEndPuncttrue
\mciteSetBstMidEndSepPunct{\mcitedefaultmidpunct}
{\mcitedefaultendpunct}{\mcitedefaultseppunct}\relax
\EndOfBibitem
\bibitem[Bene{\v s}ov{\' a} \latin{et~al.}(2025)Bene{\v s}ov{\' a}, Pokorn{\' a}, Erlebach, and Heard]{Benesova2025Mobility}
Bene{\v s}ov{\' a},~T.; Pokorn{\' a},~K.; Erlebach,~A.; Heard,~C. Mobility and {Sintering} of {Silica}-{Supported} {Platinum} {Clusters} via {Reactive} {Neural} {Network} {Potentials}. \emph{ChemRxiv} \textbf{2025}, DOI: 10.26434/chemrxiv--2025--tjz1c\relax
\mciteBstWouldAddEndPuncttrue
\mciteSetBstMidEndSepPunct{\mcitedefaultmidpunct}
{\mcitedefaultendpunct}{\mcitedefaultseppunct}\relax
\EndOfBibitem
\bibitem[Heard \latin{et~al.}(2024)Heard, Grajciar, and Erlebach]{Heard2024Migration}
Heard,~C.~J.; Grajciar,~L.; Erlebach,~A. Migration of zeolite-encapsulated subnanometre platinum clusters \textit{via} reactive neural network potentials. \emph{Nanoscale} \textbf{2024}, \emph{16}, 8108--8118\relax
\mciteBstWouldAddEndPuncttrue
\mciteSetBstMidEndSepPunct{\mcitedefaultmidpunct}
{\mcitedefaultendpunct}{\mcitedefaultseppunct}\relax
\EndOfBibitem
\bibitem[Baerlocher \latin{et~al.}()Baerlocher, Brouwer, Marler, and McCusker]{IZA}
Baerlocher,~C.; Brouwer,~D.; Marler,~B.; McCusker,~L.~B. Database of Zeolite Structures. https://www.iza-structure.org/databases/\relax
\mciteBstWouldAddEndPuncttrue
\mciteSetBstMidEndSepPunct{\mcitedefaultmidpunct}
{\mcitedefaultendpunct}{\mcitedefaultseppunct}\relax
\EndOfBibitem
\bibitem[Erlebach \latin{et~al.}(2024)Erlebach, {\v S}{\' i}pka, Saha, Nachtigall, Heard, and Grajciar]{Erlebach2024reactive}
Erlebach,~A.; {\v S}{\' i}pka,~M.; Saha,~I.; Nachtigall,~P.; Heard,~C.~J.; Grajciar,~L. A reactive neural network framework for water-loaded acidic zeolites. \emph{Nature Communications} \textbf{2024}, \emph{15}, 4215\relax
\mciteBstWouldAddEndPuncttrue
\mciteSetBstMidEndSepPunct{\mcitedefaultmidpunct}
{\mcitedefaultendpunct}{\mcitedefaultseppunct}\relax
\EndOfBibitem
\bibitem[Deng \latin{et~al.}(2023)Deng, Zhong, Jun, Riebesell, Han, Bartel, and Ceder]{deng_2023}
Deng,~B.; Zhong,~P.; Jun,~K.; Riebesell,~J.; Han,~K.; Bartel,~C.~J.; Ceder,~G. CHGNet as a pretrained universal neural network potential for charge-informed atomistic modelling. \emph{Nature Machine Intelligence} \textbf{2023}, \emph{5}, 1031--1041\relax
\mciteBstWouldAddEndPuncttrue
\mciteSetBstMidEndSepPunct{\mcitedefaultmidpunct}
{\mcitedefaultendpunct}{\mcitedefaultseppunct}\relax
\EndOfBibitem
\bibitem[Christensen and Lilienfeld(2020)Christensen, and Lilienfeld]{Christensen2020Revised}
Christensen,~A.~S.; Lilienfeld,~A.~V. Revised {MD17} dataset ({rMD17}). 2020; \url{https://figshare.com/articles/Revised_MD17_dataset_rMD17_/12672038/3}\relax
\mciteBstWouldAddEndPuncttrue
\mciteSetBstMidEndSepPunct{\mcitedefaultmidpunct}
{\mcitedefaultendpunct}{\mcitedefaultseppunct}\relax
\EndOfBibitem
\bibitem[Batatia \latin{et~al.}(2023)Batatia, Kovács, Simm, Ortner, and Csányi]{batatia2023macehigherorderequivariant}
Batatia,~I.; Kovács,~D.~P.; Simm,~G. N.~C.; Ortner,~C.; Csányi,~G. MACE: Higher Order Equivariant Message Passing Neural Networks for Fast and Accurate Force Fields. \emph{arXiv} \textbf{2023}, DOI: arXiv:2206.07697\relax
\mciteBstWouldAddEndPuncttrue
\mciteSetBstMidEndSepPunct{\mcitedefaultmidpunct}
{\mcitedefaultendpunct}{\mcitedefaultseppunct}\relax
\EndOfBibitem
\bibitem[Dodge(2008)]{Spearman}
Dodge,~Y. \emph{The {Concise} {Encyclopedia} of {Statistics}}; Springer New York, 2008; pp 502--505\relax
\mciteBstWouldAddEndPuncttrue
\mciteSetBstMidEndSepPunct{\mcitedefaultmidpunct}
{\mcitedefaultendpunct}{\mcitedefaultseppunct}\relax
\EndOfBibitem
\bibitem[Willimetz \latin{et~al.}(2025)Willimetz, Erlebach, Heard, and Grajciar]{Willimetz2025}
Willimetz,~D.; Erlebach,~A.; Heard,~C.~J.; Grajciar,~L. 27Al NMR chemical shifts in zeolite MFI via machine learning acceleration of structure sampling and shift prediction. \emph{Digital Discovery} \textbf{2025}, \emph{4}, 275--288\relax
\mciteBstWouldAddEndPuncttrue
\mciteSetBstMidEndSepPunct{\mcitedefaultmidpunct}
{\mcitedefaultendpunct}{\mcitedefaultseppunct}\relax
\EndOfBibitem
\bibitem[Willimetz \latin{et~al.}(2025)Willimetz, Mart\i{}nez-Ortigosa, Brako-Amoafo, Grajciar, Vidal-Moya, Bornes, Sarou-Kanian, Erlebach, Rey, Blasco, and Heard]{Willimetz2025Aluminum}
Willimetz,~D.; Mart\i{}nez-Ortigosa,~J.; Brako-Amoafo,~D.; Grajciar,~L.; Vidal-Moya,~A.; Bornes,~C.; Sarou-Kanian,~V.; Erlebach,~A.; Rey,~F.; Blasco,~T.; Heard,~C. Aluminum {Siting} in {Zeolite} {RTH} {From} a {Combined} {Machine} {Learning} - {NMR} {Approach}. \emph{ChemRxiv} \textbf{2025}, DOI: 10.26434/chemrxiv--2025--1p3dj\relax
\mciteBstWouldAddEndPuncttrue
\mciteSetBstMidEndSepPunct{\mcitedefaultmidpunct}
{\mcitedefaultendpunct}{\mcitedefaultseppunct}\relax
\EndOfBibitem
\bibitem[Joyce \latin{et~al.}(2007)Joyce, Yates, Pickard, and Mauri]{CASTEP}
Joyce,~S.~A.; Yates,~J.~R.; Pickard,~C.~J.; Mauri,~F. A first principles theory of nuclear magnetic resonance J-coupling in solid-state systems. \emph{The Journal of Chemical Physics} \textbf{2007}, \emph{127}, 204107\relax
\mciteBstWouldAddEndPuncttrue
\mciteSetBstMidEndSepPunct{\mcitedefaultmidpunct}
{\mcitedefaultendpunct}{\mcitedefaultseppunct}\relax
\EndOfBibitem
\end{mcitethebibliography}

\end{document}